\documentstyle[aps,epsf]{revtex}

\newcommand{\eins}{\mbox{$1 \hspace{-1.0mm}  {\bf l}$}}
\newcommand{\be}{\begin{equation}}
\newcommand{\ee}{\end{equation}}
\newcommand{\bea}{\begin{eqnarray}}
\newcommand{\eea}{\end{eqnarray}}

\newcommand{\ket}[1]{ | \, #1  \rangle}

\newcommand{\abs}[1]{ | \, #1 \,  |}

\newcommand{\mytext}[1]{\mbox{ #1}}

\begin{document}
\draft\onecolumn
\title{Optimal state estimation for $d$-dimensional quantum systems}
\author{Dagmar~Bru\ss $^1$\footnote{Present affiliation:
Inst. f\"{u}r Theoret. Physik, Universit\"{a}t Hannover, 
Appelstr. 2, D-30167 Hannover, Germany} and Chiara~Macchiavello$^{2}$}
\address{
$^1$ISI, Villa Gualino, Viale Settimio Severo 65, 10133 Torino, Italy\\
$^2$Dipartimento di Fisica ``A. Volta'' and INFM-Unit\`a di Pavia, 
Via Bassi 6, 27100 Pavia, Italy}
\date{Received \today}
\maketitle
\begin{abstract}
We establish a connection between optimal quantum cloning and optimal state 
estimation for $d$-dimensional quantum systems. 
In this way we derive an upper limit on the fidelity of 
state estimation for $d$-dimensional pure quantum states and, furthermore, 
for generalized inputs supported on the symmetric subspace.
\end{abstract}
One of the fundamental problems in quantum physics is the question of how  
well one can estimate  the state $\ket{\psi}$ of a quantum system, 
given that only a finite number of identical copies is available. 
An appropriate figure of merit in this context is the fidelity which 
will be defined below.
The optimal 
fidelity for state estimation of two-level systems has been derived in 
\cite{mapo}, and an algorithm for constructing an optimal 
positive operator valued measurement
(POVM) 
for a general  quantum system 
has been given in \cite{dbe}. 
The purpose of this letter is to derive the optimal fidelity for state 
estimation of an ensemble of $N$ identical pure
$d$-dimensional quantum systems by establishing a connection 
to optimal quantum cloning. 
\par 
In the following we will prove 
the link between optimal quantum cloning and optimal state 
estimation, using a similar line of argument as in 
\cite{bc}. 
We consider both processes to be universal,
in the sense that the corresponding
fidelity does not depend on the input state $\ket{\psi}$.
The link we want to show is given by the  equality
\begin{equation}
{{F}}_{d,est}^{opt}(N) = F _{d,QCM}^{opt}(N,\infty ) \ \ . 
\label{equal}
\end{equation}
Here $F _{d,QCM}^{opt}(N,M)$ is the fidelity of the optimal quantum 
cloner for $d$-dimensional systems, taking $N$ identical pure inputs 
and creating $M$ outputs, which was derived in \cite{Werner} to be
\begin{equation}
F_{d,QCM}^{opt}(N,M)=\frac{M-N+N(M+d)}{M(N+d)}\;\ \   . \label{Fnm}
\end{equation}
(This formula refers to the fidelity between an output one-particle reduced 
density operator and one of the identical inputs.)
  In equation (\ref{equal}) $ {{F}}_{d,est}^{opt}(N)$ is the 
optimal average 
 fidelity of state estimation for $N$ identical
 $d$-dimensional inputs, defined as
\begin{equation}
{F}_{est}=\sum_{\mu }p_{\mu }(\psi )|\langle \psi |\psi _{\mu }\rangle
|^{2} ,
\label{fid}
\ee
where $p_\mu(\psi)$ is the probability of finding outcome $\mu$ 
(to which we associate candidate $\ket{\psi_\mu}$), given that 
the inputs were in state 
$\ket{\psi}$. 
\par
Let us introduce the generalized Bloch vector $\vec{\lambda}$ 
by expanding a $d$-dimensional  density matrix in the following way
\cite{Mahler}
\be
\rho_d=\frac{1}{d} \eins +\frac{1}{2}\sum_{i=1}^{d^2-1}\lambda_i\tau_i \ \ ,
\label{pure}
\ee
where $\tau_i$ are the generators of the group $SU(d)$ with 
\be
\mytext{Tr} \ \tau_i = 0 \ ; \ \ \ \mytext{Tr}(\tau_i\tau_j) = 2 \delta_{ij} \ \ .
\ee
Note that the length of the generalized Bloch vector for pure states is
\be
\abs{\vec{\lambda}} = \sqrt{2(1-\frac{1}{d})}\ \ ,
\ee
 which reduces to the familiar case $\abs{\vec{\lambda}} = 1$
for qubits, i.e. $d=2$.
\par
It has been shown in \cite{ugly} that,
as far as optimality of the fidelity for a universal map is concerned, one can
restrict oneself to covariant transformations. Furthermore, 
 a covariant  map, 
acting on pure $d$-dimensional input states, 
 can only  shrink the generalized Bloch vector, namely 
 it transforms equation (\ref{pure}) into the output density operator
\be
\rho_d=\frac{1}{d} \eins +\frac{1}{2}\eta_d\sum_{i=1}^{d^2-1}\lambda_i\tau_i \ ,
\label{out}
\ee
where we call $\eta_d$ the shrinking factor.
Note that for pure input states the fidelity  is related to $\eta_d$ 
as follows:
\be
F_d = \frac{1}{d}[1+(d-1)\eta_d ]\ \ .
\ee
\par
Remember that, as mentioned above, in this paper we consider
universal quantum cloning and universal state estimation and therefore
the above considerations apply.
In order to clarify the role of the shrinking factor in quantum state 
estimation we notice that 
equation (\ref{fid}) can also be interpreted as
\be
{F}_{d,est}(N)=\langle \psi |{\varrho}_{d,est}|\,\psi \rangle
\ee
where the density operator ${{\varrho}_{d,est}}$, due to universality, 
is the shrunk version of the input $|\psi \rangle \langle \psi |$, namely
\begin{equation}
{\varrho}_{d,est}={\eta}_{d,est}(N)|\psi \rangle \langle \psi |+
 (1-{%
\eta}_{d,est}(N))\frac{1}{d}{\mbox{$1 \hspace{-1.0mm}  {\bf l}$}.}
\label{barrho}
\end{equation}
We now start proving the equality (\ref{equal}) by noticing that 
after performing a
 universal measurement procedure  on $N$ identically prepared 
 input copies $\ket{\psi}$, 
 we can  prepare a state of $L$ systems, supported on the symmetric
  subspace of ${\cal H}_d^{\otimes L}$, where each system has the same 
  reduced density operator, given by ${{\varrho}_{d,est}}$. The symmetric 
  subspace is defined as the space spanned by all states  which 
  are invariant under any permutation of the constituent subsystems.

\par 
As shown in \cite{Werner}, a universal cloning process generates outputs
that are supported on the symmetric subspace. Therefore, the above
method of performing state estimation followed by preparation of a
symmetric state can be viewed as a universal cloning process and thus it
cannot lead to a higher fidelity than the optimal $N\to L$ cloning
transformation. Therefore we find the inequality
\begin{equation}
{F}_{d,est}^{opt}(N)\leq F _{d,QCM}^{opt}(N,L)\ \ .
\label{leq}
\end{equation}
The above inequality
 must hold for any value of L, in particular for $L\to \infty$. 
\par In order to derive the opposite inequality,
 we consider a measurement procedure on N copies which is composed of 
 an optimal $N\to L$ cloning process and a subsequent universal measurement on 
 the $L$ output copies. 
 This total procedure is also a possible state estimation method.
 As mentioned above 
 the output $\varrho_L$ of the optimal universal $d$-dimensional cloner 
 is supported on the symmetric subspace and therefore can be decomposed as
\cite{Werner}
\be
\varrho _{L}=\sum_{i}\alpha _{i}|\psi
_{i}\rangle \langle \psi _{i}|^{\otimes L} ;  \ \ \ \  \
\mytext{with} \ \ \ \sum_{i}\alpha _{i}=1
 \ \ ,
 \label{decom}
\ee
where the coefficients $\alpha_i$ are not necessarily positive. After performing 
the optimal universal measurement on the L cloner outputs we can calculate 
the average fidelity
of the total process, due to linearity of the measurement procedure, 
as follows:
\be
{F}_{d,total}(N,L)=\sum_{i} \alpha _{i}\langle\psi |[{\eta}_{d,est}^{opt}(L)|
\psi
_{i}\rangle \langle \psi _{i}|+(1-{\eta}_{d,est}^{opt}(L))\frac{1}{d}{%
\mbox{$1 \hspace{-1.0mm}  {\bf l}$}]|}\psi \rangle \ \ .
\ee
(Remember that 
$\sum_{i}\alpha _{i}|\psi _{i}\rangle \langle \psi _{i}|$ is the one particle
reduced density matrix at the output of the
$N\to L$ cloner and thus depends on $N$ and $L$.)
In the limit $L\to \infty$ we have  ${\eta}_{d,est}^{opt}(\infty)=1$ and 
the average fidelity can be written as
\bea
\lim_{L\to\infty}{F}_{d,total}(N,L) &= &
\langle \psi |[\sum_{i}\alpha _{i}|\psi _{i}\rangle \langle \psi _{i}|]{|}\psi
\rangle \nonumber \\
& = & \langle\psi |[{\eta}_{d,QCM}^{opt}(N,\infty)|\psi
\rangle \langle \psi|+(1-{\eta}_{d,QCM}^{opt}(N,\infty))\frac{1}{d}{%
\mbox{$1 \hspace{-1.0mm}  {\bf l}$}]|}\psi \rangle  \nonumber \\
& = & \frac{1}{d}[1+(d-1)\eta _{d,QCM}^{opt}(N,\infty)]\ ,
\eea
where in the second line we have explicitly
written down the output of the cloning stage
for clarity.
This fidelity cannot be higher than the one for the optimal state estimation 
performed directly on N pure inputs, thus we conclude
\begin{equation}
{F}_{d,QCM}^{opt}(N,\infty)\leq F _{d,est}^{opt}(N)\ \ .
\label{geq}
\end{equation}
The above inequality, together with equation (\ref{leq}), leads to 
the equality (\ref{equal}).
Thus we have derived the optimal fidelity for state estimation of $N$ 
copies of a $d$-dimensional quantum system  to be
\be
{F}_{d,est}^{opt}(N) = \frac{N+1}{N+d} \ \ .
\ee
\par
Note that we can 
extend this argument for optimal state estimation to more general inputs, 
namely  to inputs supported on the symmetric subspace.
Using the decomposition (\ref{decom}), we see immediately that
 we can always reach at least the same 
shrinking factor as for pure inputs, due to linearity of the 
measurement procedure. 
Moreover, we can prove by contradiction that the shrinking factor cannot 
be larger than for pure states:
let us assume we could perform better on such a 
described entangled input. 
We can think of arranging the following procedure.
We concatenate an $N\to M$ cloning transformation taking $N$ pure inputs
and creating $M$ outputs with 
a subsequent state estimation. 
Notice that, generalizing the result of \cite{bc}, 
the shrinking factors of  two 
concatenated universal operations multiply,
given that the output of the first is supported by the symmetric subspace.
 If we could perform better than in the pure case at the second 
 stage of this concatenation, we could, 
 by reconstructing the output state according to the state estimation result,
 create an $N\to \infty$ cloner that is better than the optimal one,
 thus arriving at a contradiction.
 \par
 In conclusion, we have derived the optimal fidelity for state estimation of 
 an ensemble of identical $d$-dimensional quantum states, 
 pointing out the connection to optimal quantum cloning. We have also 
 extended the possible inputs for state estimation 
 in $d$ dimensions to
 those supported on the symmetric subspace. 
 Note that an algorithm to construct
 the according POVM consisting of a finite set of 
 operators has  been given in reference \cite{dbe}.
\par
We would like to thank A.~Ekert and
P.~Zanardi for helpful discussions. We acknowledge 
support  by the European TMR Research Network
ERP-4061PL95-1412. Part of this work was completed during the 1998 
workshops on quantum information of
ISI - Elsag-Bailey   and Benasque Center of Physics.


\begin{references}
\bibitem{mapo}  S.~Massar and S.~Popescu, Phys.~Rev.~Lett. {\bf 74}, 1259
(1995).
\bibitem{dbe} R. Derka, V. Buzek,
A. Ekert, Phys. Rev. Lett. {\bf 80}, 1571 (1998).
\bibitem{bc} D.~Bru\ss , A.~Ekert, C.~Macchiavello, 
         Phys. Rev. Lett.  {\bf 81}, 2598 (1998).
\bibitem{Werner}  R.~Werner, Phys.~Rev. A{\bf 58}, 1827 (1998).
\bibitem{Mahler} J.~Schlienz, G.~Mahler, Phys.~Rev. A{\bf 52}, 4396 (1995).
\bibitem{ugly} P.~Zanardi, Phys.~Rev. A{\bf 58}, 3484 (1998).

\end{references}
\end{document}